\documentclass[11pt]{article}

\usepackage{amsmath, amssymb, bm}
\usepackage{geometry}
\usepackage{graphicx}
\usepackage[backend=biber, style=numeric, citestyle=numeric, sorting=none,subentry=true, maxbibnames=99]{biblatex}
\usepackage{booktabs}
\usepackage[colorlinks=true,linkcolor=blue,citecolor=blue,urlcolor=blue]{hyperref}
\usepackage{xcolor}
\title{\vspace{-1.5cm}\Large \textbf{Nonequilibrium Maxwell-Demon NEMD formalism for transport}:\\[0.3em]
\textbf{I. Extrapolating shear viscosity to the hydrodynamic limit}}

\author{
Hesam Arabzadeh$^{1*}$,  Brad Lee Holian$^{2\dagger}$ \\[0.5cm]
\normalsize $^{1}$Department of Chemistry, University of Missouri, Columbia, Missouri 65211-7600, USA\\
\normalsize $^{2}$Theoretical Division, Los Alamos National Laboratory, Los Alamos, New Mexico 87545, USA\\[0.5cm]
\normalsize $^{*}$Electronic mail: \href{mailto:hacr6@missouri.edu}{hacr6@missouri.edu}\\
\normalsize $^{\dagger}$Electronic mail: \href{mailto:blhksh@gmail.com}{blhksh@gmail.com} -- Retired
}

\date{}

\begin{document}

\maketitle

\begin{abstract}
The Maxwell-Demon nonequilibrium molecular dynamics method for measuring the shear viscosity divides the fluid simulation cell into two regions of width w in the $x$-direction, with the particles allowed to move freely between the two sides. The Demon maintains equal and opposite regional average particle velocities in the $y$-direction by applying an acceleration needed to sustain the nonequilibrium steady state (NESS), which includes both total force balance and an instantaneous correction for diffusion of particles across boundaries. We show that the driven velocity profile in x is a natural hydrodynamic response and not imposed by the constraint. For the present shear geometry, the measured NESS profile in Eulerian slabs is represented accurately by a piecewise parabolic form, reminiscent of planar Poiseuille flow. The parabolic profile gives an estimate of the kinematic viscosity from work done on the shearing fluid, as well as an entropy production estimate, derived from heat removal by the Nosé--Hoover thermostat that keeps each regional average temperature constant. We show that the mechanical work supplied by the Demon equals the heat removed by the thermostat. Once NESS driving is removed, the parabolic velocity profile relaxes exponentially in short order to sinusoidal, the natural transverse momentum-diffusion eigenmode. Our Maxwell-Demon shear method provides a direct NEMD formalism for obtaining shear viscosity from momentum diffusion, work, and entropy balances. Our results in three dimensions for increasing system size $N$ (the number of particles) demonstrate that shear viscosity approaches an asymptote (the so-called hydrodynamic limit) as $N^{-1/3}$.
\end{abstract}

\section{Introduction}

Nonequilibrium molecular dynamics (NEMD) provides a direct way to measure transport coefficients by driving a system into a controlled nonequilibrium steady state (NESS) and relating the required driving, work, or heat removal to the corresponding hydrodynamic coefficient~\cite{ashurst1975dense}. In the usual equilibrium route, transport coefficients are obtained from Green--Kubo time-correlation functions~\cite{alder1970decay}. For shear viscosity, this requires integrating the pressure-tensor autocorrelation function; in practice, the running integral can converge slowly, and the plateau region may be difficult to identify unambiguously~\cite{holian1983shear}. NEMD avoids part of this difficulty by maintaining a steady nonequilibrium signal whose long-time average can be measured directly.

The sample-reservoir construction, introduced by Holian~\cite{holian2002simulations}, follows the NEMD framework for relaxation in dense molecular fluids. In that approach, a periodic-boundary-condition (PBC) computational cell is divided into two regions, called ``sample'' and ``reservoir,'' whose membership is determined by the instantaneous particle coordinates rather than by permanent labels. A NESS is maintained by thermostatting or by mechanical driving, and the corresponding relaxation rate is inferred from the steady-state bookkeeping. Holian showed that, when the steady-state signal is sufficiently large, the externally driven NEMD approach can be more accurate than direct adiabatic relaxation. The same paper also gave thermal-conduction and shear-flow examples, showing that the sample-reservoir idea is not restricted to vibrational relaxation.

For example, in the heat-flow case, we imagine that a Maxwell Demon maintains two regions at different temperatures under PBCs while particles move freely between the two sides. The temperature-control mechanism then provides the heat-flow bookkeeping and, through the entropy balance, reveals the asymmetry between heating and cooling. On the other hand, the same Maxwell-Demon idea can be applied to shear flow: instead of maintaining two different temperatures, the Demon maintains opposite average flow velocities in the $y$-direction on the two sides of the cell in the $x$-direction.

The present paper is the first in a series on Maxwell-Demon NEMD for hydrodynamic transport coefficients. Here we focus on shear viscosity for a Lennard--Jones (LJ) fluid. The periodic simulation box is divided into two $x$-regions of width $w$. The Demon maintains opposite regional average velocities: $\langle u_y\rangle=-u_p$ for $x<0$ and $\langle u_y\rangle=+u_p$ for $x>0$. The resulting steady shear state is analyzed using Eulerian velocity profiles, with particles assigned to slabs of width $\Delta x$ in the $x$-direction according to their instantaneous $x$-coordinate. A reasonable assumption is that the shape of the velocity profile in the periodic box of length $L_x=2w$ is sinusoidal, the natural hydrodynamic momentum-diffusion reference mode of the longest wavelength. At NESS, equal and opposite Demon accelerations maintain the imposed regional velocities against viscous relaxation, giving the transverse momentum-relaxation rate and therefore the kinematic shear viscosity. The main purpose of this paper is to derive and test the shear viscosity obtained from three equivalent routes: momentum diffusion, work, and entropy. The work route follows from the rate of mechanical work required to maintain the two regional velocities, while the entropy route follows from the heat removed from the driven steady state by the Nos\'e--Hoover thermostat, which keeps the average temperature constant in both regions ~\cite{nose1984molecular, hoover1985canonical,evans1985nose,arabzadeh2025chaos}. In a spatially varying flow field, we show that the rate of viscous dissipation is determined by the spatial average of the squared strain rate, $\langle\dot\epsilon_{xy}^{2}\rangle$, with a geometric factor connecting the driving, dissipation, and relaxation that depends on the measured velocity profile. We show that the Demon-driven NESS is consistently described by a parabolic-profile geometry, reminiscent of plane Poiseuille flow, while the sinusoidal representation is the appropriate relaxation eigenmode only when the NESS driving is turned off. The momentum-diffusion, work, and entropy expressions give consistent estimates of $\nu$ for the parabolic NESS profile.

The remainder of the paper is organized as follows. Section~\ref{sec:theory} describes the Maxwell-Demon shear geometry, the equations of motion, and the Nos\'e--Hoover temperature-control scheme, and derives the momentum-diffusion, work, and entropy routes to the kinematic viscosity. Section~\ref{sec:simulation} describes the simulation protocol, reduced units, Maxwell-Demon driving, and Eulerian data analysis. Section~\ref{sec:results} presents the results, beginning with the equilibrium equation-of-state benchmark, followed by the measured velocity, strain-rate, density, temperature, and shear-stress profiles, relaxation from the NESS, the strain-rate dependence of the viscosity, and the hydrodynamic-limit extrapolation and finite-size scaling. Section~\ref{sec:conclusions} summarizes the main conclusions and future extensions of the Maxwell-Demon framework to other transport coefficients.

\section{Theoretical formalism for shear flow by a Maxwell Demon}
\label{sec:theory}
\subsection{Maxwell-Demon shear geometry and Nosé--Hoover equations of motion}
\subsubsection{Periodic two-region geometry}

We consider an LJ fluid in a periodic simulation cell of length $L_x=2w$ in the $x$-direction. The cell is divided into two spatial regions, $-w \leq x < 0$ and $0 \leq x < w$. Originally, these left and right regions were labeled ``sample'' and ``reservoir''~\cite{holian2002simulations}. The assignment of a particle to either region is made from its instantaneous $x$-coordinate. Thus the two regions are Eulerian, not permanent Lagrangian groups of particles. A particle can leave one region and enter the other during the dynamics, with diffusion occurring along the regional boundaries.

To impose a shear state, the Maxwell Demon maintains opposite regional average velocities in the $y$-direction on the two sides of the cell. We write this condition in terms of the total particle velocity as
\begin{equation}
\langle \dot y\rangle_-=-u_p, \qquad \langle \dot y\rangle_+=+u_p ,
\end{equation}
where the subscripts $-$ and $+$ denote the regions $x<0$ and $x>0$, respectively. The instantaneous particle velocity is the peculiar (thermal) velocity plus the imposed regional center-of-mass velocity,
\begin{equation}
\dot{\mathbf r}_i=\mathbf u_i\pm u_p\hat{\mathbf y},\qquad i\in\pm.
\end{equation}
The peculiar velocity $\mathbf u_i$ is the velocity relative to the imposed average motion of the region containing particle $i$. The regional velocity constraint is equivalently written as a condition on the peculiar momentum, with $\mathbf p_i=m_i\mathbf u_i$ (particle mass is $m_i$). For each region,
\begin{equation}
\sum_{i\in\pm}m_i u_{iy}=0.
\label{eq:peculiar-momentum}
\end{equation}
Therefore the average $y$-velocity (CM, or center of mass) of each region (total mass $M_{\pm}$) is fixed:
\begin{equation}
\frac{1}{M_{\pm}}\sum_{i\in\pm}m_i\dot y_i=\pm u_p,\qquad M_{\pm}=\sum_{i\in\pm}m_i.
\label{eq:tot_M}
\end{equation}

\subsubsection{Nosé--Hoover equations of motion for Maxwell-Demon NEMD}

For a particle in either $\pm$ region, the Maxwell-Demon dynamics is written compactly as follows. From this point, the particle index $i$ and regional $\pm$ flags are suppressed unless they are needed explicitly. The signs of $u_p$ (the fixed fluid velocity) and $g_0$ (the acceleration applied in the same direction, on average) are understood from the region ($\pm$):
\begin{equation}
\begin{aligned}
\dot{\mathbf r} &= \mathbf u+u_p\hat{\mathbf y},\\
m\dot{\mathbf u} &= \mathbf F+mg_0\hat{\mathbf y}-\xi m\mathbf u,\\
\dot{\xi} &= \frac{\nu_T^2}{T_0}(T-T_0).
\end{aligned}
\label{eq:eom-pos}
\end{equation}
The inter-particle force is obtained from the total potential energy, $\Phi$:
\begin{equation}
\mathbf F=-\frac{\partial\Phi\{\mathbf r\}}{\partial\mathbf r}.
\end{equation}
The fluctuating Demon acceleration, $g_0$, is obtained from the regional force balance:
\begin{equation}
Mg_0=-\sum F_y.
\end{equation}
Particle diffusion in the boundary layers at the edges of the two regions requires $\Delta g$, an additional instantaneous acceleration, to be added to the total, $g=g_0+\Delta g$, which is used in evaluating the rate of mechanical work applied to the shearing fluid. The temperature in each region is fixed by the deterministic Nos\'e--Hoover thermostat~\cite{nose1984molecular,hoover1985canonical}. The two regions have separate NH thermostatting variables, $\xi_{\pm}$, but both thermostats have the same target temperature $T_0$. The peculiar kinetic energy, $K$, determines the regional temperature, $T$:
\begin{equation}
K\{\mathbf p\} = \sum\frac{|\mathbf p|^2}{2m} = \frac{1}{2}\sum m|\mathbf u|^2 = \frac{3}{2}Nk_B T.
\label{eq:kinetic}
\end{equation}
The instantaneous regional temperatures fluctuate, but in the NESS the Nosé--Hoover feedback maintains $\langle T_-\rangle = \langle T_+\rangle = T_0$ within statistical uncertainty.

\subsubsection{Demon acceleration from the regional peculiar-momentum constraint}

The Demon acceleration is fixed by the requirement that the regional average velocities remain equal to $\pm u_p$. Since the imposed motion is in the $y$-direction, the relevant constraint is the vanishing of the regional peculiar momentum in the $y$-direction, Eq.~\ref{eq:peculiar-momentum}. The time derivative of the total peculiar momentum gives
\begin{equation}
0=\sum m\dot u_y=\sum F_y+Mg_0-\xi\sum m u_y .
\end{equation}
The last term vanishes due to the peculiar-momentum constraint in each region, so that the continuous Demon acceleration is
\begin{equation}
g_0=-\frac{1}{M}\sum F_y .
\label{eq:demon-accel}
\end{equation}

The acceleration is not prescribed as a constant external field. Instead, it fluctuates in time so that the regional average velocities remain fixed. At the NESS, the two accelerations are equal and opposite within statistical uncertainty.

\subsubsection{Central-difference peculiar-velocity correction}

The peculiar-velocity correction is applied separately in each region. At each central-difference time step $\Delta t$,
\begin{equation}
\dot y\left(t-\frac{\Delta t}{2}\right)
=
u_y\left(t-\frac{\Delta t}{2}\right)+u_p .
\end{equation}
\begin{equation}
y(t)=y(t-\Delta t)+\dot y\left(t-\frac{\Delta t}{2}\right)\Delta t+O(\Delta t^3).
\end{equation}
The total regional acceleration includes force balance, along with a small correction due to diffusion across boundary edges in $\Delta t$:
\begin{equation}
g(t)=g_0(t)+\Delta g(t),
\end{equation}
so that the peculiar-velocity update is
\begin{equation}
u'_y\left(t+\frac{\Delta t}{2}\right)
=
u_y\left(t-\frac{\Delta t}{2}\right)
+\dot u'_y(t)\Delta t+O(\Delta t^3),
\qquad
m\dot u'_y(t)=F_y(t)+mg(t).
\end{equation}
Summing the acceleration over either region gives
\begin{equation}
\sum m\dot u'_y(t)
=
\left[\sum F_y(t)+Mg_0(t)\right]
+M\Delta g(t)
=
M\Delta g(t).
\end{equation}
The peculiar-velocity correction $\Delta u$ is defined by
\begin{equation}
M\Delta u=\sum m u'_y\left(t+\frac{\Delta t}{2}\right).
\end{equation}
Because $\sum m u_y(t-\Delta t/2)=0$,
\begin{equation}
M\Delta u
=
\sum m u_y\left(t-\frac{\Delta t}{2}\right)
+\Delta t\sum m\dot u'_y(t)
=
M\Delta g(t)\Delta t.
\end{equation}
The corrected peculiar velocities are
\begin{equation}
u_y\left(t+\frac{\Delta t}{2}\right)
=
u'_y\left(t+\frac{\Delta t}{2}\right)-\Delta u ,
\end{equation}
which guarantees
\begin{equation}
\sum m u_y\left(t+\frac{\Delta t}{2}\right)=0.
\end{equation}
Therefore,
\begin{equation}
\Delta g(t)=\frac{\Delta u}{\Delta t}.
\end{equation}

When the regional signs are written explicitly, the long-time scalar correction is
\begin{equation}
\langle\Delta g\rangle
=
\frac{1}{2}
\left(
\langle\Delta g_+\rangle-\langle\Delta g_-\rangle
\right).
\end{equation}
\subsection{Energy balance, work, and entropy}

Linear hydrodynamic theory -- the Navier--Stokes laws of momentum flux plus Fourier's law of heat flow -- assumes local thermodynamic equilibrium, where equation of state and transport properties depend on local density and temperature. For shear flow, the momentum flux $P_{xy}$ (shear pressure, or negative shear stress) is linearly related to the gradient of the fluid velocity:
\begin{equation}
P_{xy}=-\eta\dot\epsilon_{xy}
=-\eta\frac{d}{dx}u_y,
\end{equation}
where $\eta$ is the shear viscosity. The atomistic expression for the shear pressure is:
\begin{equation}
P_{xy}=\frac{1}{V}\sum_i\left[m u_{xi}u_{yi}+\frac{1}{2}\sum_{j\ne i}x_{ij}F_{y,ij}\right].
\label{eq:shear_stress}
\end{equation}

\subsubsection{Regional energy balance}

The Maxwell Demon atomistic method determines shear viscosity from the long-time average energy balance in each driven region at the nonequilibrium steady state (NESS):
\begin{equation}
\langle\dot E\rangle
=
\langle\dot Q\rangle-\langle\dot W\rangle
=0,
\qquad
\langle\dot Q\rangle=\langle\dot W\rangle<0,
\label{eq:regional-first-law}
\end{equation}
where $\langle\dot Q\rangle$ is the rate of heat added to the interior of the region, and $\langle\dot W\rangle$ is the rate of work done by the region on the outside world. With this convention, heat removal from the region corresponds to $\langle\dot Q\rangle<0$, and work done on the region by the Demon corresponds to $\langle\dot W\rangle<0$.

For each of the two regions, the total energy is
\begin{equation}
E
=
K\{\mathbf p\}
+\Phi\{\mathbf r\}
+\frac{3}{2}Nk_BT_0\left(\frac{\xi}{\nu_T}\right)^2
+\frac{1}{2}Mu_p^2,
\label{eq:regional-energy}
\end{equation}
i.e., atomistic kinetic and potential energies, plus a harmonic energetic penalty for the ministrations of the NH thermostat variable, and finally, the kinetic energy associated with the imposed regional center-of-mass velocity.

\begin{align}
\dot E
&=
\sum\left(
m\dot{\mathbf u}\cdot\mathbf u
+\frac{\partial\Phi}{\partial\mathbf r}\cdot\dot{\mathbf r}
\right)
+3Nk_BT_0\frac{\xi\dot\xi}{\nu_T^2}
\nonumber\\
&=
\sum\left[
\left(\mathbf F+mg_0\hat{\mathbf y}-\xi m\mathbf u\right)\cdot\mathbf u
-\mathbf F\cdot\left(\mathbf u+u_p\hat{\mathbf y}\right)
\right]
+3Nk_B(T-T_0)\xi
\nonumber\\
&=
g_0\sum m u_y
-u_p\sum F_y
-\xi\sum m|\mathbf u|^2
+3Nk_B(T-T_0)\xi
\nonumber\\
&=
Mg_0u_p-3Nk_BT_0\xi .
\label{eq:combined-energy-rate}
\end{align}

At the NESS, we substitute
$\langle g\rangle=\langle g_0\rangle+\langle\Delta g\rangle$
for $g_0$ in the above rate of energy change in order to account for the small difference due to diffusion along regional boundaries. With the momentum decay rate $\gamma$ defined by
$\langle g\rangle=\gamma u_p$, we have the power injected into the fluid:
\begin{equation}
\langle\dot W\rangle
=
\left\langle P_{xy}V\dot\epsilon_{xy}\right\rangle
=
-\eta V\left\langle\dot\epsilon_{xy}^2\right\rangle
=
-\gamma Mu_p^2<0.
\label{eq:workdot}
\end{equation}

Heat is continually extracted from shearing fluid due to work done on it:
\begin{equation}
\langle\dot Q\rangle
=
T_0\langle\dot S\rangle
=
-3Nk_BT_0\langle\xi\rangle
\quad\Rightarrow\quad
\langle\dot S\rangle
=
-3Nk_B\langle\xi\rangle<0
\qquad
(\langle\xi\rangle>0).
\label{eq:entropy-heat-rate}
\end{equation}
Entropy $S$ continually decreases with time, thanks to the positive NH thermostat variable acting upon particle momenta, on average, like a friction term. The phase-space distribution at NESS collapses onto a strange attractor of dimension $<6N+1$ ($S\to-\infty$).

\noindent
The shear viscosity is then determined in two expressions: work and entropy (mass density is mass $M$ per volume $V$: $\rho=M/V=Nm/V$):
\begin{equation}
\eta
=
-\frac{\langle\dot W\rangle}
{V\left\langle\dot\epsilon_{xy}^2\right\rangle}
=
\frac{M}{V}
\frac{\gamma u_p^2}
{\left\langle\dot\epsilon_{xy}^2\right\rangle}
=
\rho
\frac{\gamma u_p^2}
{\left\langle\dot\epsilon_{xy}^2\right\rangle}
\qquad\text{(``work'')}.
\end{equation}
\begin{equation}
\eta
=
-\frac{\langle\dot Q\rangle}
{V\left\langle\dot\epsilon_{xy}^2\right\rangle}
=
\frac{3Nk_BT_0\langle\xi\rangle}
{V\left\langle\dot\epsilon_{xy}^2\right\rangle}
=
\rho
\frac{3k_BT_0\langle\xi\rangle}
{m\left\langle\dot\epsilon_{xy}^2\right\rangle}
\qquad\text{(``entropy'')}.
\end{equation}

The kinematic viscosity (momentum diffusivity) is
\begin{equation}
\nu=\frac{\eta}{\rho}=\frac{\lambda^2}{\tau},
\end{equation}
where $\lambda$ = mean free path, $\tau=\eta/B$ = collision time
($B=\rho c^2$ = bulk modulus, $c$ = bulk sound speed):
\begin{equation}
\nu
=
\frac{\gamma u_p^2}
{\left\langle\dot\epsilon_{xy}^2\right\rangle}
=
\frac{3k_BT_0}{m}
\frac{\langle\xi\rangle}
{\left\langle\dot\epsilon_{xy}^2\right\rangle}
\qquad\text{(``work'' vs.\ ``entropy'')}.
\end{equation}

\subsection{Hydrodynamic representations and momentum-diffusion estimate}

\subsubsection{Parabolic NESS profile}

The Maxwell-Demon constraint fixes the regional averages, but it does not prescribe the local velocity profile within either half of the cell. The profile shape must therefore be measured from the Eulerian velocity field. We find that the driven NESS profile is accurately represented by a piecewise parabolic form, rather than by a pure sinusoidal mode. For $x>0$ $(+)$:
\begin{equation}
u_y(x)
=
6u_p\frac{x}{w}\left(1-\frac{x}{w}\right),
\qquad
\langle u_y\rangle
=
\frac{1}{w}\int_0^w dx\,u_y(x)
=
u_p .
\end{equation}
\begin{equation}
\frac{du_y}{dx}
=
\frac{6u_p}{w}
\left(1-2\frac{x}{w}\right)
=
\dot\epsilon_{xy},
\qquad
\left\langle\dot\epsilon_{xy}^2\right\rangle
=
12\frac{u_p^2}{w^2}.
\label{eq:parabolic-strain-factor}
\end{equation}
\begin{equation}
\frac{d^2u_y}{dx^2}
=
-12\frac{u_p}{w^2}
=
\left\langle\frac{d^2u_y}{dx^2}\right\rangle.
\end{equation}
\begin{equation}
\nu_{\mathrm{work}}
=
\frac{\gamma w^2}{12}
=
\frac{k_BT_0}{4mu_p^2}\langle\xi\rangle w^2.
\end{equation}

\subsubsection{Relaxation of the parabolic regional average}

The parabolic profile is not a single eigenfunction of the momentum-diffusion operator. Therefore its full relaxation is not described by one exact exponential mode in the same way as the sinusoidal profile. However, the initial decay rate of the regional average is obtained directly from the same momentum-diffusion equation:
\begin{equation}
\frac{d\langle u_y\rangle}{dt}
=
\nu\frac{d^2\langle u_y\rangle}{dx^2},
\qquad
\nu=\frac{\eta}{\rho}
=
\text{kinematic viscosity}.
\end{equation}

The sign of the curvature of the NESS velocity profile in each region ($\pm$) is opposite to the sign of the long-time spatial average of $u_y(x,t<0)$.

For exponential relaxation ($x>0$):
\begin{equation}
\langle u_y\rangle(t>0)
=
\langle u_y\rangle(0)e^{-\gamma t}
=
u_p e^{-\gamma t},
\end{equation}
\begin{equation}
\frac{d\langle u_y\rangle}{dt}(t>0)
=
-\gamma u_p e^{-\gamma t}
\quad\Rightarrow\quad
\frac{d\langle u_y\rangle}{dt}(0)
=
-\gamma u_p
=
\nu\frac{d^2\langle u_y\rangle}{dx^2}(0).
\end{equation}

We can then evaluate the kinematic viscosity from the initial ($t=0$) relaxation from the NESS established for $-\infty<t<0$:
\begin{equation}
\nu_{\mathrm{relax}}
=
-\frac{\gamma\langle u_y\rangle}
{\left\langle d^2u_y/dx^2\right\rangle}.
\end{equation}
For the parabolic profile,
\begin{equation}
\nu_{\mathrm{relax}}
=
\frac{\gamma w^2}{12}
=
\nu_{\mathrm{work}}.
\end{equation}

The sinusoidal expression for the shearing velocity profile is presented in \ref{app:sinusoidal-reference}. It remains useful as the corresponding eigenmode estimate for relaxation to equilibrium once the NESS external driving has been stopped. The NESS velocity profile is parabolic, not sinusoidal, a fact that was unappreciated in earlier work.~\cite{holian2002simulations}

\section{Simulation protocol and data analysis}
\label{sec:simulation}

\subsection{Lennard-Jones fluid and reduced units}

The representative Maxwell-Demon shear simulation was carried out for an LJ fluid with $N=1872$ particles. All quantities are reported in reduced LJ units using the convention $r_0=m=\epsilon=1$, where the length unit $r_0$ is the pair separation at the LJ potential minimum, $m$ is the atomic mass, $\epsilon$ is the LJ potential well depth, and the resulting unit of time is $t_0=r_0\sqrt{m/\epsilon}$; $r_0=2^{1/6}\sigma$ is the position of the LJ minimum (for historical reasons, $\sigma$ is the repulsive hard-sphere crossing point of the LJ potential). In these $r_0=1$ units the unsmoothed LJ potential is
\begin{equation}
\phi(r)=r^{-12}-2r^{-6}, \label{eq:lj}
\end{equation}
so that the potential minimum is located at $r=1$ and has depth $-1$. We smoothed the potential beyond the minimum, matching both pair potential and force at the inflection point $r_c$ with a cubic-spline function, and ending with zero potential and force at the cutoff $r_m$ (see ~\ref{app:cubic-spline-lj}).

The target temperature was $T_0=2.75$, and the mass density (in $r_0=1$ units) was $\rho=0.99$. This choice for the equilibrium state of the LJ fluid corresponds to twice the critical temperature and halfway between critical and triple-point densities; therefore, the fluid is sufficiently far from both the liquid--gas coexistence dome and the freezing line in the LJ phase diagram.~\cite{holian1983shear} Our nonequilibrium study focuses on the weak shear-thinning regime, approaching the zero-strain-rate (Green--Kubo) limit. The fluid is far from being turbulent and remains (as is expected for Navier--Stokes--Fourier hydrodynamics) in local thermodynamic equilibrium (LTE), i.e., nowhere near the strong shock regime in dense fluids, where small deviations from LTE have been observed.~\cite{holian2024shockwave} The box length in the gradient direction was $L_x=2w$, where both the $\pm$ regional widths were $w=10.35$; the transverse dimensions were $L_y=L_z=9.55$.

The equations of motion were integrated using St{\o}rmer (``leapfrog'') finite centered differences, as derived and tested in our previous work.~\cite{holian1995thermostatted, arabzadeh2026heating} The time step was chosen to be somewhat conservative, i.e., $dt=0.001$, and the production simulations (long-time NESS averages) were run for $10^8$ time steps, or $10^5t_0$. Measurements were accumulated following a time of $50t_0$ for thermal equilibration (including rapid melting of the initial face-centered cubic or simple cubic crystal).

\subsection{Maxwell-Demon driving and Nosé--Hoover parameters}

The Maxwell-Demon shear simulations used the two-region geometry described above. At each step, particles were assigned to the left or right region according to the sign of their instantaneous $x$-coordinate. The Demon acceleration was then chosen separately for the two regions to maintain the imposed regional average velocities. The regional peculiar-velocity correction was applied after every central-difference leapfrog velocity update. The two regions were thermostatted separately with Nosé--Hoover heat-flow variables, $\xi_+$ and $\xi_-$, but both were coupled to the same target temperature, $T_0=2.75$. The NH thermostat rate parameter was set equal to the mean collision rate, $\nu_T=10.36$, which was determined by counting the number of times the instantaneous kinetic temperature crossed its running mean in an equilibrium NVE simulation, as described in our previous work.~\cite{holian1995thermostatted, arabzadeh2026heating} (Note: $\nu_T$ has no direct connection to, and should not be confused with, the kinematic viscosity $\nu$.)

\subsection{Eulerian slab profiles}

Eulerian slab profiles were accumulated after initial equilibration using $40$ bins in the $x$-direction. Slab quantities (velocity components, density, temperature, energy, pressure components, shear stress, kinetic stress contributions, and heat flux) were sampled every $10^4$ time steps ($10t_0$). These quantities were averaged over long times to construct profiles. The strain-rate profile was obtained from the slab-averaged $y$-velocity profile either by finite differences between neighboring Eulerian slabs or by differentiating the fitted profile. The velocity profile was compared with both the sinusoidal reference form and a piecewise parabolic profile. The parabolic fit was used to evaluate the main profile-dependent geometric factor, while the sinusoidal fit was retained only as a reference eigenmode comparison. The shear stress profile was computed from the kinetic and configurational contributions, Eq.~\ref{eq:shear_stress}, with the sums evaluated for the particles and pair contributions assigned to the Eulerian slab.

\section{Results}
\label{sec:results}

\subsection{Equilibrium equation-of-state benchmark}

Before considering the nonequilibrium shear response, we benchmarked the equilibrium thermodynamic properties of the LJ fluid against the equation-of-state representation used by Holian \textit{et al.} \cite{holian1980shock}. For this comparison, we used the standard LJ 12--6 potential rather than the spline-modified interaction employed in the shear simulations, since the historical equation of state corresponds to the standard LJ fluid. A simulation with $N=500$ atoms was performed at $\rho=1$ and $T=2.75$, using a cutoff of $4\sigma$ with analytical long-range corrections. The pressure, internal energy, isothermal bulk modulus, and constant-volume heat capacity are compared in Table~\ref{tab:lj_eos}. The historical values were obtained by evaluating the analytical Helmholtz-free-energy representation in Appendix A of Ref.~\cite{holian1980shock} at the same thermodynamic state.

\begin{table*}[t]
\centering
\caption{Equilibrium thermodynamic properties of the LJ fluid at $\rho r_0^3=1$ and $k_{\mathrm B}T/\epsilon=2.75$. Present results are from an $N=500$ LJ simulation; historical values are from the \ref{app:sinusoidal-reference}  fit of Holian \textit{et al.} \cite{holian1980shock}.}
\label{tab:lj_eos}

\begin{tabular}{lcccc}
\toprule
& \multicolumn{1}{c}{Present work}
& \multicolumn{1}{c}{Holian \textit{et al.}} \\
\cmidrule(lr){2-3}\cmidrule(lr){4-5}
Property & $r_0$ units & $r_0$ units \\
\midrule
$P^*$     & $7.3210\pm0.0010$ & $7.3333$ \\
$E^*/N$  & $0.1706\pm0.0002$ & $0.1316$ \\
$B_T^*$ & $24.178\pm0.0340$ & $24.4308$ \\
$C_V/(Nk_{\mathrm B})$ & $1.9943\pm0.0012$ & $1.9841$ \\
\bottomrule
\end{tabular}

\vspace{3pt}
\begin{minipage}{0.98\textwidth}
\footnotesize
Here $P_\ell^*=P\ell^3/\epsilon$ and $B_{T,\ell}^*=B_T\ell^3/\epsilon$, with $\ell=r_0$. $E^*/N$ and $C_V/(Nk_{\mathrm B})$ are independent of the length convention.
%Here $P_\ell^*=P\ell^3/\epsilon$ and $B_{T,\ell}^*=B_T\ell^3/\epsilon$, with $\ell=r_0$ or $\sigma$. Since $r_0=2^{1/6}\sigma$, $P_{r_0}^*=\sqrt{2}\,P_\sigma^*$ and $B_{T,r_0}^*=\sqrt{2}\,B_{T,\sigma}^*$. $E^*/N$ and $C_V/(Nk_{\mathrm B})$ are independent of the length convention.
\end{minipage}
\end{table*}

The present pressure agrees with the historical fit to within approximately $0.2\%$, while $B_T$ and $C_V$ differ by approximately $1.0\%$ and $0.5\%$, respectively. The internal energy shows a larger relative difference, with $E/N=0.1706$ compared with $0.1316$ from the historical fit. Overall, the comparison provides an independent equilibrium benchmark of the thermodynamic state used to characterize the LJ fluid before considering its nonequilibrium transport response.

\subsection{Velocity profile: sinusoidal reference and parabolic NESS fit}

\begin{figure}
\centering
\includegraphics[width=0.45\linewidth]{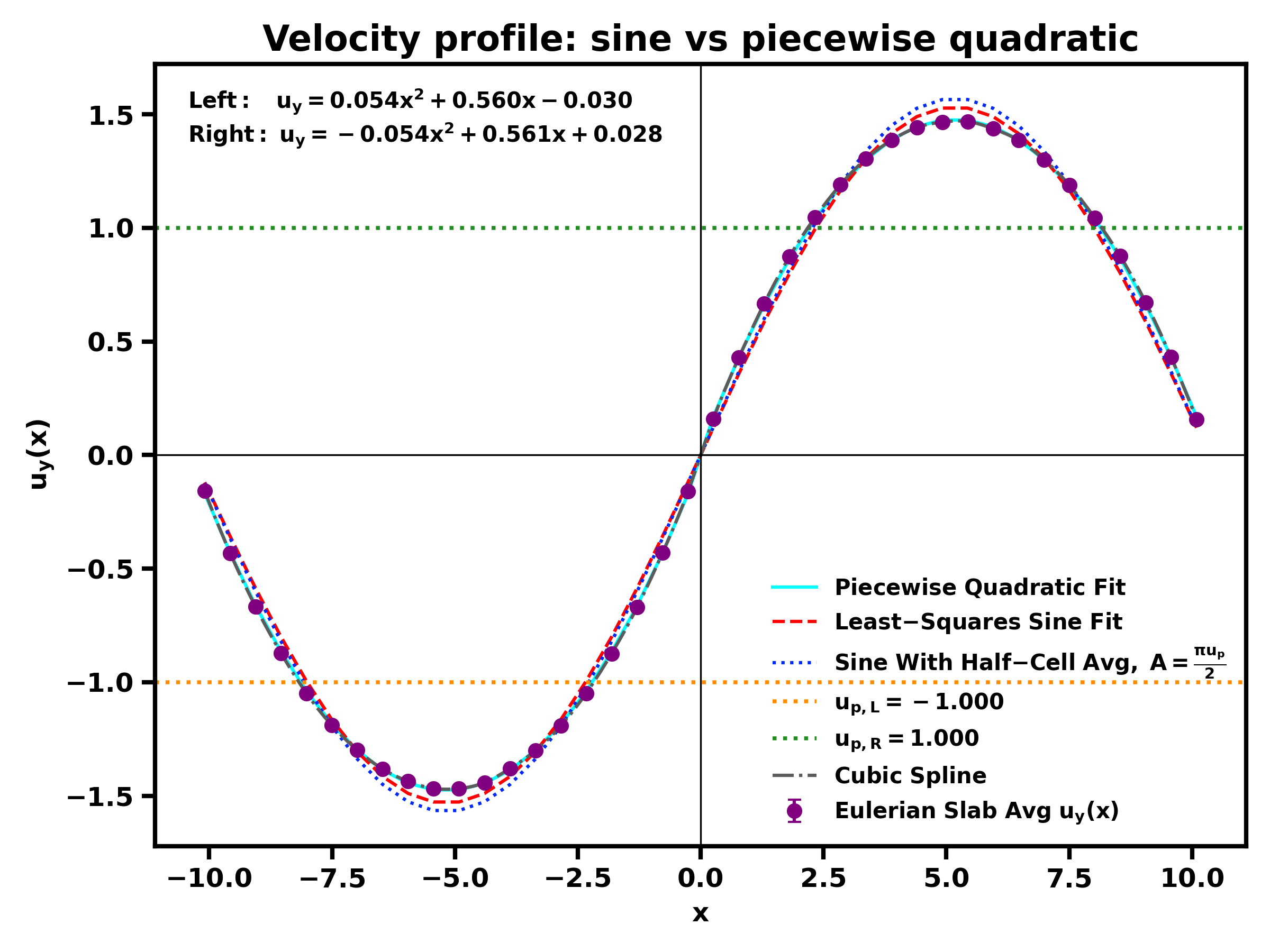}
\caption{Time-averaged velocity profile $u_y$ for the representative Maxwell-Demon shear simulation. The imposed regional averages are $\langle u_y\rangle=\pm u_p$, shown by the dotted horizontal lines. Eulerian slab averages are compared with a cubic-spline interpolation, a least-squares sinusoidal fit, and a piecewise quadratic fit on the left and right halves of the cell. Both functional forms describe the data well, but the piecewise quadratic fit provides the more accurate representation of the steady NESS profile.}
\label{fig:uy-profile}
\end{figure}

Figure~\ref{fig:uy-profile} shows the resulting time-averaged velocity profile for the representative run with input $u_p=1$. The measured regional averages are very close to the target values. The measured Demon-driven NESS profile is accurately represented by a piecewise parabolic form on the two half-cells. The horizontal dotted lines in Fig.~\ref{fig:uy-profile} mark the imposed regional averages $\pm u_p$. The comparison shows that the Demon constraint does not impose either sinusoidal or parabolic functional form directly; rather, it imposes the two regional means, and the measured Eulerian profile determines the parabolic geometric factor used in the viscosity estimate.

\subsection{Demon acceleration and steady-state profiles}

\begin{figure}
\centering
\includegraphics[width=0.45\linewidth]{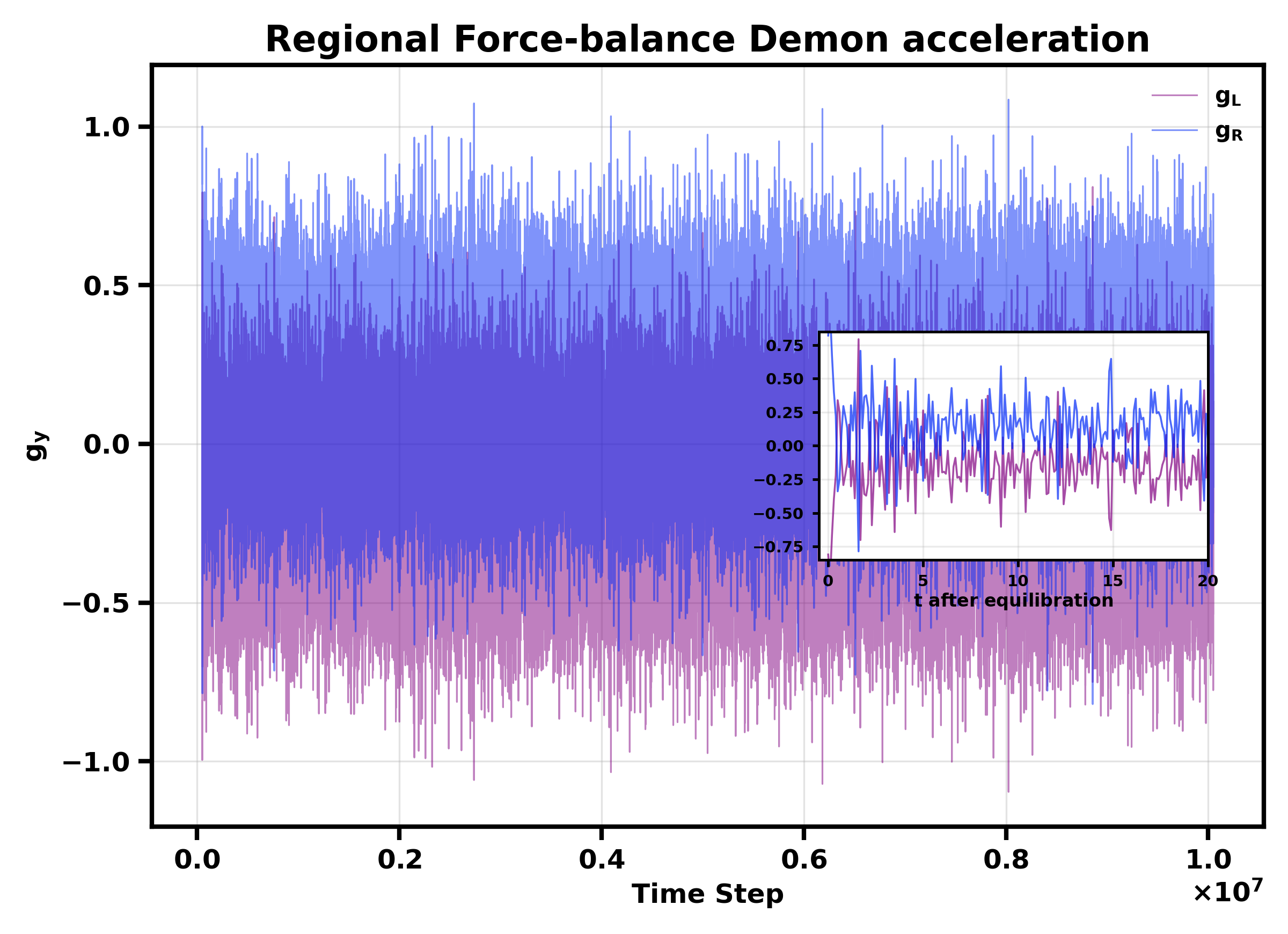}
\caption{Time series of the regional force-balance Demon accelerations applied to the two halves of the periodic cell. These plotted accelerations correspond to the continuous force-balance contribution, $g_0$. They fluctuate as they balance instantaneous force fluctuations in each region, but their long-time averages are equal and opposite. The inset shows the same acceleration fluctuations over the first 20 reduced time units after equilibration. For the representative run, $g_0=0.148$; $\Delta g=0.036$. The mechanical work balance uses the total acceleration $g=g_0+\Delta g=0.184$.}
\label{fig:g-timeseries}
\end{figure}

Figure~\ref{fig:g-timeseries} shows the time series of the two regional force-balance accelerations for the representative run. The accelerations fluctuate strongly because they compensate the instantaneous force and momentum exchange between the two regions. Their long-time averages are equal and opposite within statistical uncertainty. For the representative run, $\langle g\rangle=\pm0.184$ and $u_p=0.998$, giving $\gamma=\langle g\rangle/u_p=0.184$. This total momentum relaxation rate is used in the parabolic work and momentum-diffusion viscosity estimates.

\begin{figure}
\centering
\includegraphics[width=0.45\linewidth]{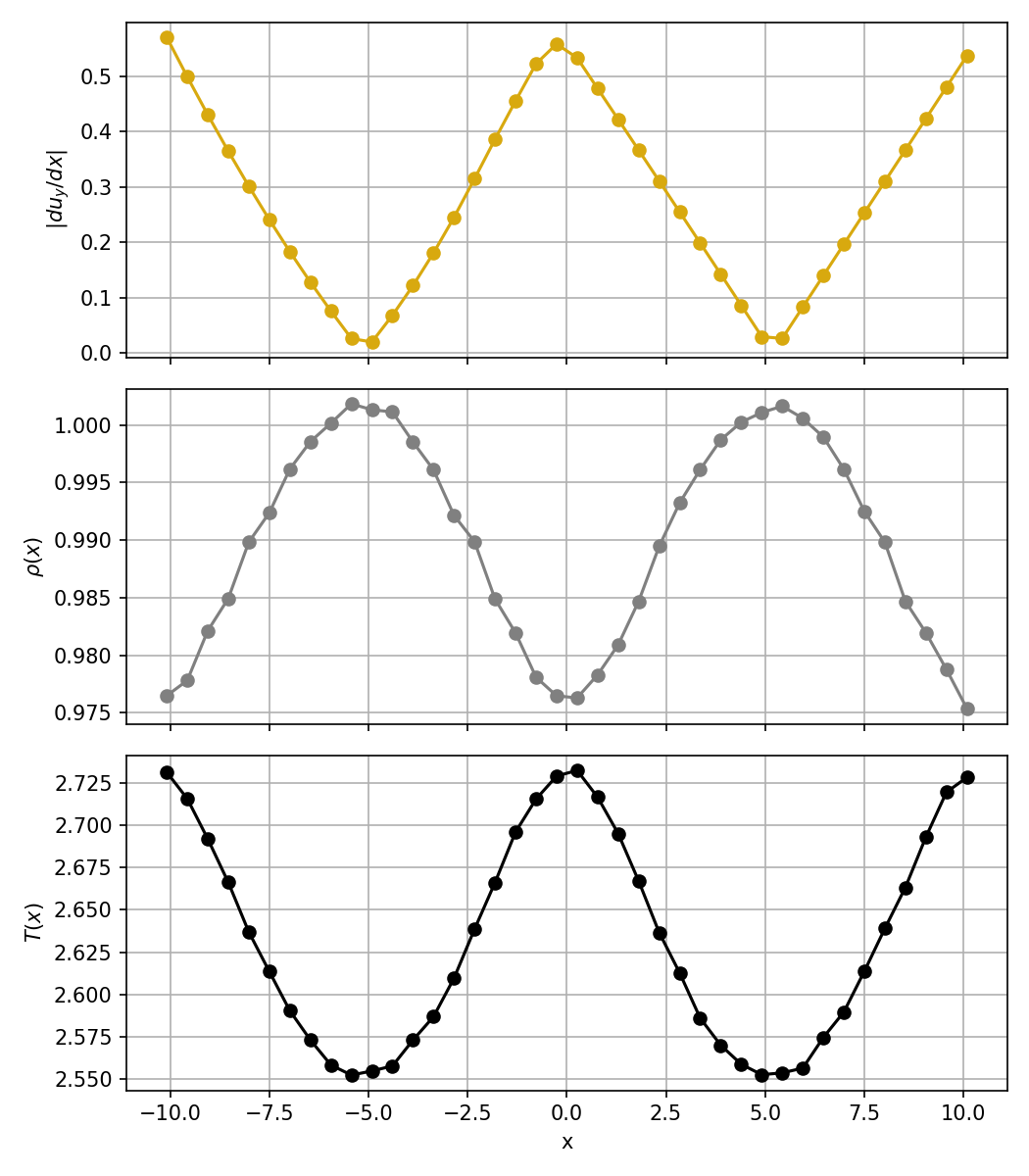}
\caption{Eulerian slab profiles for the representative Maxwell-Demon shear simulation. The upper panel shows the magnitude of the local strain rate obtained from the velocity profile. The measured strain-rate profile follows the parabolic NESS geometry used in the viscosity estimation. The middle and lower panels show the Eulerian density and local temperature profiles, respectively.}
\label{fig:eps-rho-T}
\end{figure}

The corresponding Eulerian slab profiles are shown in Fig.~\ref{fig:eps-rho-T}. The upper panel gives the magnitude of the local strain rate obtained from the derivative of the time-averaged velocity profile. For the parabolic NESS representation, the maximum strain-rate magnitude is $\dot\epsilon_{xy}^{\max}=6u_p/w=0.578$. The corresponding spatial average of the squared strain rate is $0.111$, which is used for the work and entropy estimates. The middle and lower panels of Fig.~\ref{fig:eps-rho-T} show the density and temperature profiles indicating that the Maxwell-Demon acceleration maintains the shear flow with very small density or temperature modulations. The heat generated by the driven shear flow is removed by the Nosé--Hoover feedback, leaving a stationary shear NESS.

\subsection{Shear-pressure profile}

A further test of the steady shear state is provided by the Eulerian profile of the shear pressure. In the hydrodynamic description, the local shear pressure is related to the local strain rate by the Navier--Stokes relation,
\begin{equation}
P_{xy}(x)=-\eta\frac{du_y}{dx}. \label{eq:shear}
\end{equation}

\begin{figure}
\centering
\includegraphics[width=0.45\linewidth]{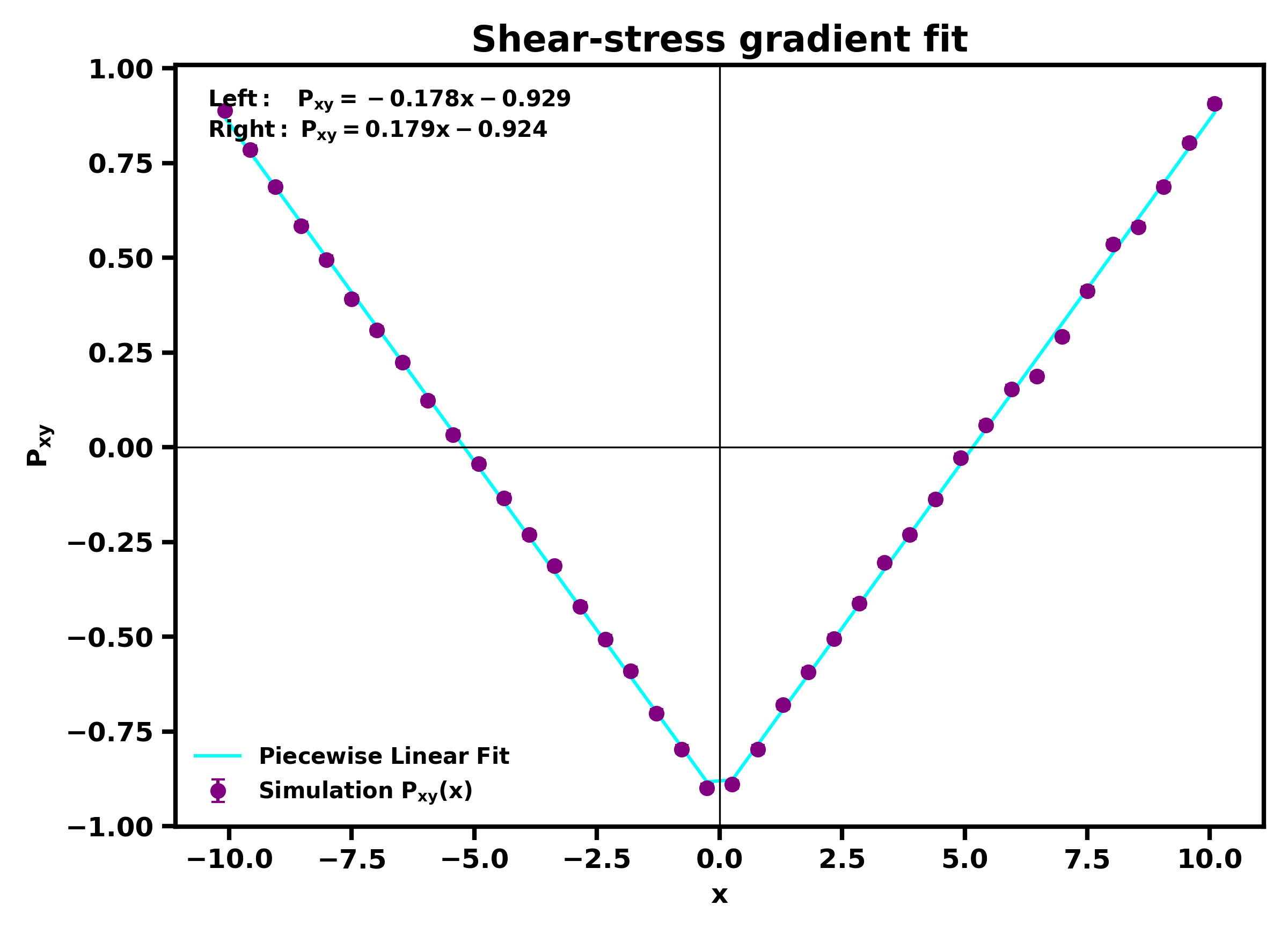}
\caption{Time-averaged Eulerian profile of the shear pressure $P_{xy}(x)$ for the representative Maxwell-Demon shear simulation. The data are compared with piecewise linear fits on the two half-cells. For the parabolic NESS velocity profile, $du_y/dx$ is linear within each half-cell, so the Navier--Stokes constitutive relation $P_{xy}(x)=-\eta du_y/dx$ predicts an approximately piecewise linear shear-pressure profile.}
\label{fig:shear-stress-profile}
\end{figure}

Thus the shear pressure profile should follow the spatial structure of $-du_y/dx$. For the parabolic NESS profile, the strain rate is linear within each half-cell.
Therefore the shear pressure is expected to vary approximately linearly within each half-cell, with the sign determined by the local velocity gradient. Figure~\ref{fig:shear-stress-profile} shows the slab-averaged profile of $P_{xy}(x)$ for the representative run together with piecewise linear fits on the two half-cells. The pressure profile is smooth and has the expected symmetry: it is negative where $du_y/dx>0$ and positive where $du_y/dx<0$. The fitted slopes have nearly equal magnitude and opposite sign, consistent with the parabolic NESS geometry. Thus the measured pressure profile provides an independent Eulerian check on the constitutive relation used in the viscosity analysis.

The pressure-gradient form gives an additional profile-based viscosity diagnostic:
\begin{equation}
\nu_{\mathrm{prof}} = -\frac{1}{\rho}\frac{\frac{dP_{xy}}{dx}}{\frac{d^2u_y}{dx^2}}.
\end{equation}
For the parabolic profile, $d^2u_y/dx^2$ is constant within each half-cell. Therefore the slope of the measured pressure profile can be compared directly with the curvature of the measured velocity profile. This provides a profile-based check on the viscosity obtained from the Demon work and entropy balances.
\subsection{Relaxation from the NESS}

The relaxation calculation provides an independent test of the momentum-diffusion interpretation. In the driven simulation, the Maxwell Demon maintains the regional average velocities for $t<0$ by applying the regional accelerations $\mathbf g_{\pm}$. At $t=0$, the driving acceleration is removed. The regional average velocities are then no longer constrained, and the remaining velocity profile relaxes toward equilibrium.

\begin{figure}
\centering
\includegraphics[width=0.45\linewidth]{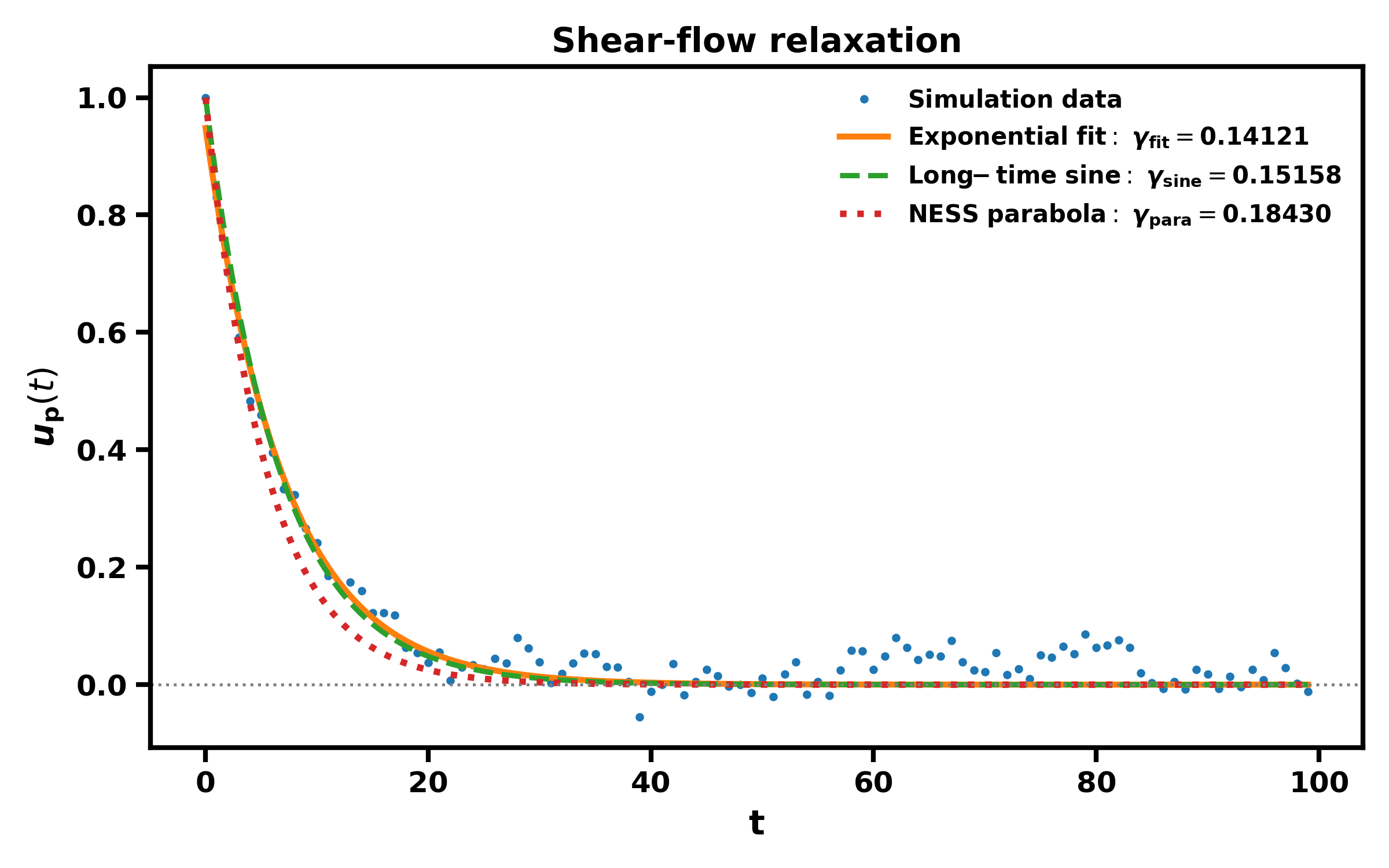}
\caption{Exponential relaxation of the shear velocity amplitude after the Maxwell-Demon NESS driving acceleration is removed at $t=0$. The initial value is $\gamma_{\mathrm{para}}(0)=\langle g\rangle/u_p=0.1843$; the predicted sinusoidal eigenmode relaxation rate is $\gamma_{\mathrm{sine}}(0)=0.1516=(\pi^2/12)\gamma_{\mathrm{para}}(0)$, compared to the observed fit, $\gamma_{\mathrm{obs}}(t\to\infty)=0.1412$, confirming transverse momentum diffusion.}
\label{fig:relaxation-fit}
\end{figure}

Figure~\ref{fig:relaxation-fit} shows the decay of the shear amplitude after the Demon acceleration is removed. The measured relaxation is described by a single exponential over the fitting interval, giving the relaxation rate $\gamma$. This confirms that the long-time decay is controlled by transverse momentum diffusion and that the sinusoidal mode provides the appropriate eigenmode description of the late-stage relaxation.

\subsection{Kinematic viscosity as a function of strain rate}

The preceding sections focused on a representative Maxwell-Demon shear simulation. We now consider the dependence of the measured kinematic viscosity on the imposed strain-rate amplitude. For the observed parabolic NESS profile, the maximum strain rate is
\begin{equation}
\dot\epsilon_{xy}^{\max}=\frac{6u_p}{w}.
\end{equation}

For each driving amplitude, the regional velocity amplitude $u_p$ and total driving acceleration $g$ were measured after initial equilibration of the fluid. At NESS, the heat removed by the Nosé--Hoover thermostat independently balances the work supplied by the Demon.
\begin{figure}
\centering
\includegraphics[width=0.45\linewidth]{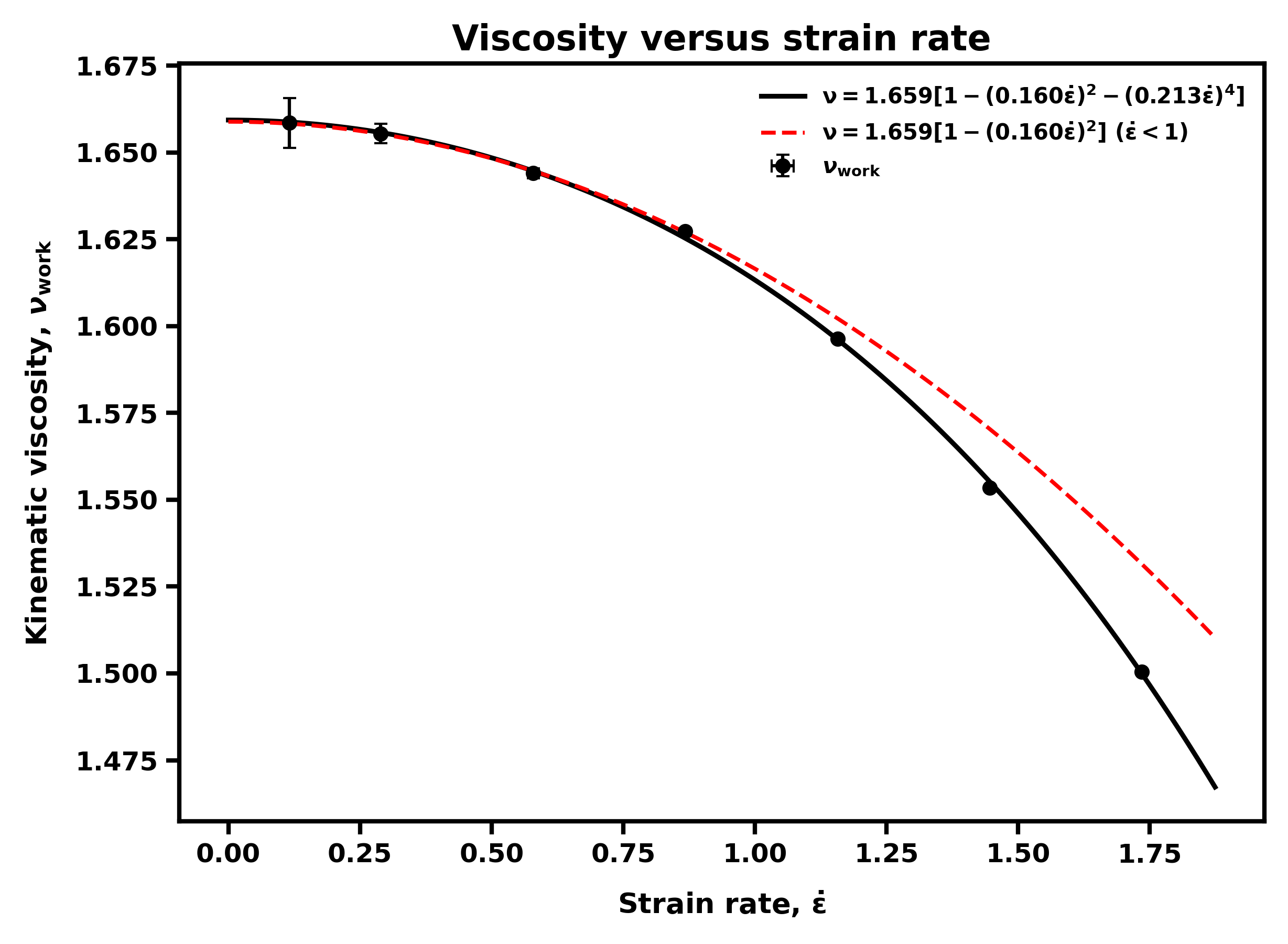}
\caption{Kinematic viscosity from Maxwell-Demon NEMD as a function of the parabolic maximum strain rate $\dot\epsilon_{xy}^{\max}=6u_p/w$. The viscosity is evaluated from the parabolic work and regional-average relaxation expression $\nu_{\mathrm{para}}=\gamma w^2/12$. The quadratic shear-thinning time, $\tau_2=0.160$, is very close to values obtained earlier by Holian~\cite{holian2002simulations} ($0.161$) and Holian and Evans~\cite{holian1983shear} ($0.153$).}
\label{fig:nu-strainrate}
\end{figure}

\begin{table}[t]
\centering
\small
\caption{Parabolic maximum strain rate and kinematic viscosity obtained from the mechanical-work and thermostat-entropy routes for each imposed regional velocity amplitude (note that the long-time average of the fluid velocity differs slightly from the imposed value, $u_p$). The entropy route gives consistent, though slightly higher, viscosities over the imposed strain-rate range.}
\label{tab:nu-strainrate}
\begin{tabular}{rrrr}\\
\hline
$u_p$ & $\dot\epsilon_{xy}^{\max}$ & $\nu_{\mathrm{work}}$ & $\nu_{\mathrm{entropy}}$ \\
\hline
0.2 &  0.1157 &  1.6584 &  1.6818 \\
0.5 &  0.2893 &  1.6554 &  1.6659 \\
1.0 &  0.5786 &  1.6440 &  1.6540 \\  
1.5 &  0.8679 &  1.6272 &  1.6348 \\  
2.0 &  1.1571 &  1.5963 &  1.6052 \\   
2.5 &  1.4462 &  1.5535 &  1.5636 \\   
3.0 &  1.7352 &  1.5005 &  1.5111 \\   
\hline
\end{tabular}
\end{table}
Figure~\ref{fig:nu-strainrate} shows the resulting kinematic viscosity as a function of $\dot\epsilon_{xy}^{\max}$, and the numerical values are listed in Table~\ref{tab:nu-strainrate}. The parabolic work and entropy routes give consistent viscosities over the imposed strain-rate range. At low strain rate, the viscosity approaches the Green--Kubo limit. At larger imposed strain rates, the viscosity of the LJ fluid decreases, demonstrating so-called "shear thinning.'' The black curve, essentially passing through all the NEMD viscosities, is a sum of quadratic plus quartic terms in strain rate. The dashed red curve is the quadratic approximation, fitted to data below a strain rate of unity. For the representative run discussed above, the parabolic work and regional-average relaxation route gives $\nu_{\mathrm{para}}=1.644$; the entropy route gives $1.654$, differing by less than $0.6\%$. 

\subsection{Hydrodynamic-limit extrapolation and finite-size scaling}
\label{sec:hydrodynamic_limit}

Molecular-dynamics simulations of fluids at equilibrium employ periodic boundary conditions (PBC) that mimic an infinite system without discontinuous boundary layers caused by free surfaces or hard containment walls. The usual worldwide geometry is a checkerboard of identical replicates of the computational cell, where an atom that leaves the box in a given time step re-enters as its image on the opposite side with the same momentum. Equilibrium is described by the long-time average of the global density and temperature, and the equation of state (EOS) properties (pressure and energy) demonstrate system-size dependence on the total volume of the computational cell containing $N$ atoms; EOS properties per particle then approach the thermodynamic limit at infinite system size like $1/N$.

Under nonequilibrium conditions, however, the fluid is no longer a homogeneous isotropic system, but rather displays disturbances in local density, temperature, and fluid velocity -- the so-called hydrodynamic fields. For nonequilibrium molecular-dynamics simulations of the hydrodynamic transport coefficients, inhomogeneous disturbances in the fields give a more pronounced dependence upon $N\to\infty$ than at equilibrium -- the so-called hydrodynamic limit, which also depends upon the dimensionality ($d=1$, 2, or 3) of the PBC computational box.

Whenever a collision occurs between two particles in a computational cell, a sound wave propagates outward at the bulk sound speed $c_B$ until the signal reaches the periodic boundary and begins to cancel with its images, resulting in a sonic cutoff of low frequencies ($0<\omega<\omega_{\min}$) and wavelengths greater than $L$, the length of the periodic cell. The bulk sound-wave traversal time of the periodic cell is
\begin{equation}
t_L=L/c_B\sim N^{1/d}\quad\Rightarrow\quad\omega_{\min}\sim 1/t_L\sim N^{-1/d}.
\end{equation}
In 3D, the sonic correction due to system size therefore goes like $N^{-1/3}$. Alternatively, the disturbance can be modeled by a diffusive Gaussian ``wave'' whose origin is at the collision point in space and time. When the sound wave first reaches the PBC boundary, the signal inside the box has diffusively spread in proportion to the square root of time (whether in 1D, 2D, or 3D): the width of the diffusive ``wave'' is proportional to $\sqrt{t_L}$, so that its volume in $d$ dimensions is $t_L^{d/2}\sim(N^{1/d})^{d/2}\sim\sqrt{N}$. The amplitude of the initial collisional impulse is reduced by the inverse volume of the diffusional wave ($t_L^{-d/2}$) analogous to the long-time tail of the velocity autocorrelation function.\footnote{The self-diffusion coefficient depends on the time integral of the velocity autocorrelation function. In 3D, the sonic correction is proportional to $1/\sqrt{t_L}\sim N^{-1/6}$, a very slow convergence with system size; however, in 2D, the self-diffusion coefficient diverges with system size like $\ln(N)$.~\cite{holian2021boundary}} Cancellation with its own periodic images means that the diffusional correction due to system size goes like $1/\sqrt{N}$, much like the fluctuations from the central limit theorem, regardless of dimensionality. The question for the hydrodynamic limit is: in 3D, does it go like the sonic model ($N^{-1/3}$) or the diffusive model ($N^{-1/2}$)? Note that for two dimensions, the two models are indistinguishable.

In the present shear geometry, transverse momentum is transported along the $x$-direction over the full periodic wavelength $L_x=2w$. The relevant hydrodynamic finite-size correction may therefore be governed primarily by this longitudinal dimension rather than by the total particle number alone. To examine the system-size dependence of the measured shear viscosity, we performed two finite-size sequences at fixed density and fixed imposed regional velocity $u_p$. In the cubic sequence, the particle number was increased from $N=512$ to $64{,}000$, with $L_x=L_y=L_z$, so that all three box dimensions increased together. In the non-cubic (or ``extended'' PBC cell-shape) sequence, the particle number was increased from $N=935$ to $2804$, the transverse dimensions were held fixed at $L_y=L_z=9.55$, and only the full box length $L_x$ in the shear-gradient direction was increased. In both sequences, the half-width entering the parabolic solution is $w=L_x/2$. For the cubic systems, $w = L_x/2 = L_y/2 = L_z/2$ whereas for the extended systems only $L_x$ varies. Because the finite-size calculations were performed at fixed $u_p$, the maximum parabolic strain rate $\dot\epsilon_{xy}^{\max}=6u_p/w$ decreases as the system size increases. To separate finite-strain-rate shear thinning from the system-size dependence, we used the quadratic+quartic constants $\tau_2 = 0.160$ and $\tau_4 = 0.213$ from Fig.~\ref{fig:nu-strainrate} to extrapolate strain rate-dependent values of $\nu$ to zero strain rate (Green-Kubo) values, $\nu(0)$, as given in Table~\ref{tab:finite-size-viscosity}.

\begin{figure}
\centering
\includegraphics[width=0.45\textwidth]{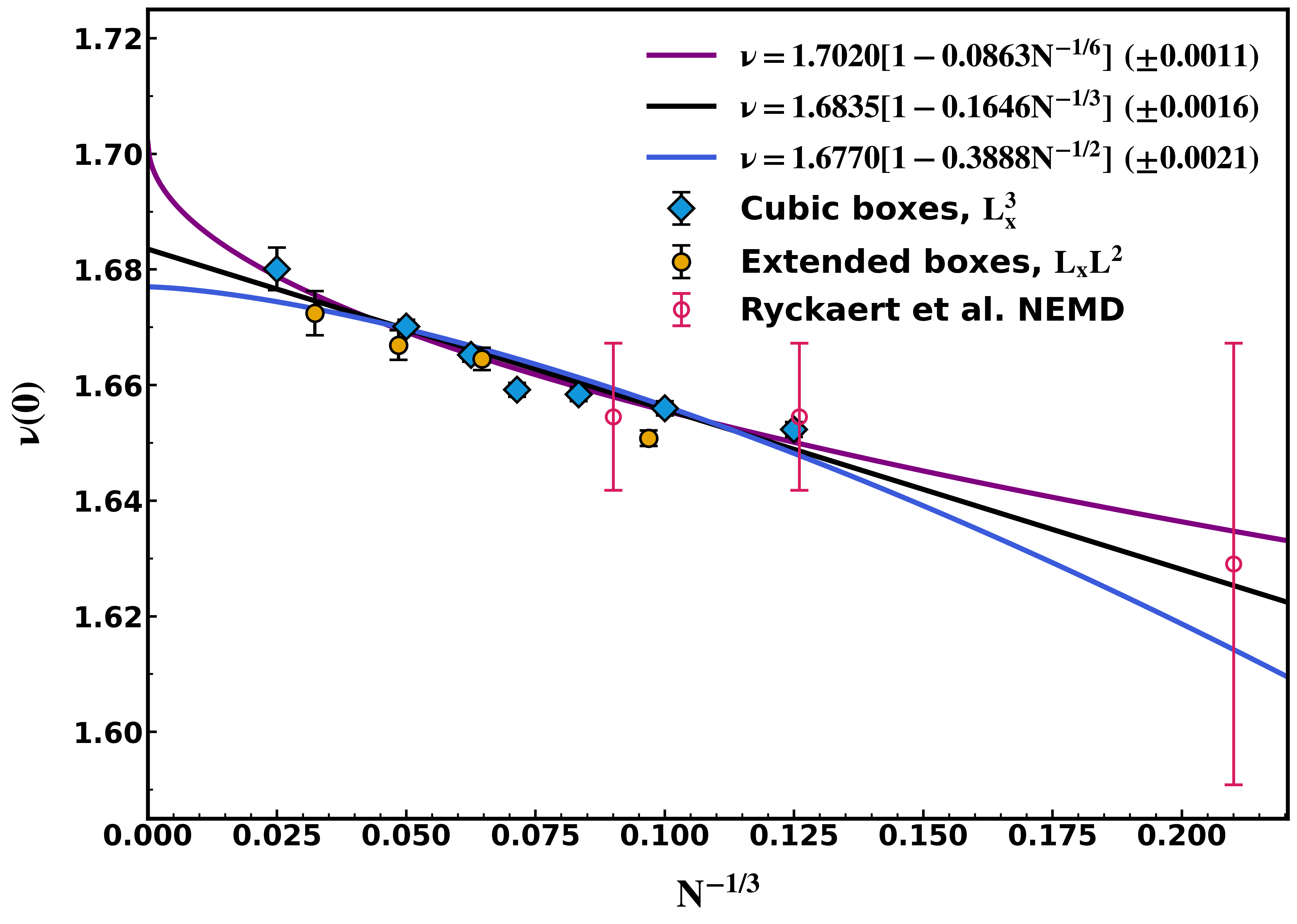}
\caption{Finite-size dependence of the zero-strain-rate kinematic shear viscosity $\nu(0)$ as a function of $N^{-1/3}$. Nonlinear least-squares fits are shown for $N^{-1/6}$ (red curve), $N^{-1/3}$ (black straight line), and $N^{-1/2}$ (blue curve), along with relative root mean-square errors in parentheses next to the fits. Error bars on $\nu(0)$ data points denote the propagated standard errors, all less than $0.4\%$. Blue diamonds are cubic-shaped ($L_x^3$) computational-cell results used in the fitting procedures; orange circles denote non-cubic ($L_xL^2$, $N=\rho L_x^3/m$) systems. Magenta open circles are from NEMD simulations by Ryckaert \textit{et al.}~\cite{ryckaert1989evaluation}; their Green--Kubo results are not shown, since they display much larger error bars.}
\label{fig:hydrodynamic_limit}
\end{figure}

\begin{table*}
\centering
\small
\caption{Finite-size dependence of the zero-strain-rate kinematic shear viscosity,
$\nu(0)=\nu(\dot\epsilon_{xy}^{\max})/
[1-(0.160\dot\epsilon_{xy}^{\max})^2-(0.213\dot\epsilon_{xy}^{\max})^4]$.}

\label{tab:finite-size-viscosity}
\begin{tabular}{lrrrrrrrr}\\
\hline
Cell geometry & $N$ & $L_x$ & $L_y$ & $L_z$ & $\dot\epsilon_{xy}^{\max}$ & $\nu(\dot\epsilon_{xy}^{\max})$ & $\nu(0)$ & $\sigma_{\nu(0)}$ \\
\hline
Cubic & 512   & 8.0268   & 8.0268   & 8.0268   & 1.480042 & 1.543330 & 1.652306 & 0.001277 \\
Cubic & 1000 & 10.0336 & 10.0336 & 10.0336 & 1.184022 & 1.589819 & 1.655948 & 0.001234 \\
Cubic & 1728 & 12.0403 & 12.0403 & 12.0403 & 0.986686 & 1.613841 & 1.658409 & 0.001210 \\
Cubic & 2744 & 14.0470 & 14.0470 & 14.0470 & 0.845732 & 1.627058 & 1.659186 & 0.001198 \\
Cubic & 4096 & 16.0537 & 16.0537 & 16.0537 & 0.740016 & 1.640873 & 1.665246 & 0.001190 \\
Cubic & 8000 & 20.0671 & 20.0671 & 20.0671 & 0.592014 & 1.654728 & 1.670135 & 0.001180 \\
Cubic & 64000 & 40.1342 & 40.1342 & 40.1342 & 0.296007 & 1.676291 & 1.680086 & 0.003697 \\
Extended & 1099 & 10.3555 & 9.5500 & 9.5500 & 1.147216 & 1.589290 & 1.650795 & 0.001333 \\
Extended & 3706 & 15.5277 & 9.5500 & 9.5500 & 0.765084 & 1.638404 & 1.664521 & 0.001937 \\
Extended & 8781 & 20.6999 & 9.5500 & 9.5500 & 0.573916 & 1.652490 & 1.666918 & 0.002553 \\
Extended & 29651 & 31.0553 & 9.5500 & 9.5500 & 0.382543 & 1.666100 & 1.672439 & 0.003800 \\
\hline
\end{tabular}
\end{table*}

The dimensions, kinematic viscosities, and their standard errors are summarized in Table~\ref{tab:finite-size-viscosity} as a function of system size. Figure~\ref{fig:hydrodynamic_limit} shows the parabolic kinematic viscosity, extrapolated to zero strain rate, $\nu(0)$, as a function of system size $N$ for the cubic-shaped cells. The data for the non-cubic cell shape show no significant systematic disagreement, which is consistent with the interpretation that the dominant finite-size effect is controlled primarily by the longitudinal dimension $L_x=2w$, while the fixed transverse dimensions of the extended systems have a weaker influence on the shear response. We are therefore justified to report the effective number of atoms for a non-cubic system as $N=\rho L_x^3/m$ in Table~\ref{tab:finite-size-viscosity}, and the viscosities are shown by the non-cubic symbols in Fig.~\ref{fig:hydrodynamic_limit}. For the cubic cell simulations, nonlinear least-squares fits to three finite-size scaling models -- sonic ($N^{-1/3}$), diffusional ($N^{-1/2}$), and long-time-tail model for self-diffusion ($N^{-1/6}$) -- are shown. All three models are barely distinguishable for intermediate system sizes; divergence from the error bars is more pronounced at both the largest and smallest sizes. The diffusional model diverges most noticeably, while the long-time-tail model appears to match the data best. The gentleness of the hydrodynamic limit as $N\to\infty$ makes a firm declaration difficult, but we can rule out $N^{-1/2}$.

Earlier comparisons of equilibrium Green--Kubo and NEMD shear viscosities examined Green--Kubo systems containing $N=108$ to $4000$ particles but used only $N=108$ for the NEMD calculation.~\cite{holian1983shear} Within the statistical uncertainty available at that time, no clear system-size dependence was resolved, and the small-system NEMD result was assumed to represent the infinite-system limit. In 1988, the first proof of the analyticity of shear thinning of viscosity with strain rate, namely, downward quadratic behavior, was demonstrated by Ryckaert et al.~\cite{ryckaert1998shear}. In subsequent NEMD shearing simulations, Ferrario et al.~\cite{ferrario1991shear} confirmed this quadratic analyticity; however, due to large error bars for the four system sizes they studied, it isn't possible for us to determine, unambiguously, the functional form of the hydrodynamic limit from their data. The present Maxwell-Demon simulations have small statistical uncertainties that allow us to see a modest but systematic finite-size dependence. Two-dimensional NEMD simulations of shear viscosity by Hoover and Posch demonstrated diffusional dependence ($1/N^{1/2}$) on system size~\cite{hoover1995shear}, but no discrimination between sonic versus diffusional models is possible, since they are identical in 2D. Here, in 3D, the sonic $1/L\sim1/N^{1/3}$ dependence shows a better goodness-of-fit than the diffusive (fluctuation) model. However, somewhat to our surprise, we cannot rule out the long-time-tail model for self-diffusion ($1/N^{1/6}$), since its goodness-of-fit is the best of the three models.

\section{Conclusions}
\label{sec:conclusions}

We developed a Maxwell-Demon formalism for measuring shear viscosity at the atomistic scale, and tested it on a fluid of atoms interacting by the well-known Lennard-Jones potential. The method imposes equal and opposite regional average velocities in two halves of a periodic boundary condition simulation cell. Particles are not permanently assigned to either region; instead, their membership is determined by their instantaneous position. The Demon acceleration required to maintain the regional velocity constraint provides the mechanical driving needed to balance viscous momentum relaxation.

A central result is that the Demon constrains only the regional average velocities, not the pointwise velocity field. The local steady-state profile must therefore be measured from Eulerian slab averages. Although the sinusoidal profile is the natural transverse momentum-diffusion eigenmode, the driven NESS profile is better represented by a piecewise parabolic form on the two half-cells. The parabolic representation gives an estimate of viscosity $\nu_{\mathrm{para}}=\gamma w^2/12$, where $\gamma=g/u_p$, and $g=g_0+\Delta g$ (force balance and diffusion correction to the total acceleration).

The work and entropy routes give consistent viscosities when evaluated with the same parabolic strain-rate factor, $\left\langle\dot\epsilon_{xy}^2\right\rangle=12u_p^2/w^2$ [Eq.~\ref{eq:parabolic-strain-factor}]. In the representative simulation, the parabolic work estimate and the Nosé--Hoover entropy estimate differ by less than $0.6\%$. This agreement shows that the mechanical work supplied by the Maxwell Demon is balanced by the heat removed by the thermostat at NESS. The shear-pressure profile provides an additional check: because the parabolic velocity profile has a linear strain-rate profile within each half-cell, the measured shear pressure is also approximately piecewise linear, consistent with linear Navier--Stokes hydrodynamics, $P_{xy}=-\eta du_y/dx$.

The sinusoidal representation is the late-time relaxation eigenmode after the driving is removed, confirming transverse momentum diffusion. At early times, the value of $\gamma$ obtained from the NESS parabolic profile describes the exponential relaxation rate.

Overall, the Maxwell-Demon shear method provides a compact nonequilibrium route to shear viscosity based on regional momentum control, mechanical work, and thermostat heat removal. The method also highlights the importance of using the measured Eulerian profile shape when converting Demon driving and dissipation into a viscosity.

Finally, the approach to infinite system size with the number of particles $N$ (the so-called hydrodynamic limit) is rather weak for shear viscosity in moderately dense fluids in three dimensions. (The entire range for the observed viscosities is only about $3\%$. Standard errors for the viscosities were always kept below about $0.4\%$, with nonlinear least-squares RMS errors no more than half that.) We have shown that the limit appears to go like $1/L\sim1/N^{1/3}$, though, at least on grounds of its smaller nonlinear least-squares error, $1/N^{1/6}$ behavior (reminiscent of the long-time tail for self-diffusion) cannot be ruled out.

Future applications can extend the same Maxwell-Demon framework to other transport coefficients, including thermal conductivity and longitudinal viscosity.

\section*{Data Availability}
The data that support the findings of this study are available within the article. All results were generated by direct numerical simulations using equations and methods fully specified in the manuscript.

\appendix
\renewcommand{\thesection}{Appendix~\Alph{section}}
\section{Sinusoidal reference profile and relaxation}
\label{app:sinusoidal-reference}

A useful reference representation of the time-averaged shear profile is the sinusoidal momentum-diffusion mode,
\begin{equation}
u_y(x)=\frac{\pi}{2}u_p\sin\left(\frac{\pi x}{w}\right), \qquad \langle u_y\rangle=\frac{1}{w}\int_0^w dx\,u_y(x)=u_p .
\end{equation}
\begin{equation}
\frac{du_y}{dx} = \frac{\pi^2u_p}{2w}\cos\left(\frac{\pi x}{w}\right) = \dot\epsilon_{xy}, \qquad\left\langle\dot\epsilon_{xy}^2\right\rangle = \frac{\pi^4}{8}\frac{u_p^2}{w^2}.
\end{equation}
\begin{equation}
\frac{d^2u_y}{dx^2} = -\frac{\pi^2}{w^2}u_y(x), \qquad \left\langle\frac{d^2u_y}{dx^2}\right\rangle = -\frac{\pi^2}{w^2}\langle u_y\rangle.
\end{equation}

Thus, the sinusoidal profile approximation would yield the following expression for the work-related kinematic viscosity:
\begin{equation}
\nu_{\mathrm{work}}
= \frac{\gamma w^2}{\pi^4/8} = \frac{k_BT_0}{(\pi^4/24)mu_p^2}\langle\xi\rangle w^2, \qquad \frac{\pi^4}{8}=12.18,
\end{equation}
while the relaxation expression gives:
\begin{equation}
\nu_{\mathrm{relax}} = -\frac{\gamma\langle u_y\rangle} {\left\langle d^2u_y/dx^2\right\rangle} = \frac{\gamma w^2}{\pi^2} = \frac{\pi^2}{8}\nu_{\mathrm{work}} = 1.234\,\nu_{\mathrm{work}}.
\end{equation}

From the work expression, the sinusoidal approximation appears to nearly mimic the observed parabolic profile; however, the relaxation result is noticeably different:
\begin{equation}
\frac{\nu_{\mathrm{work}}^{\mathrm{sinusoidal}}} {\nu_{\mathrm{work}}^{\mathrm{parabolic}}} = \frac{12}{\pi^4/8}
= 0.986, \qquad \frac{\nu_{\mathrm{relax}}^{\mathrm{sinusoidal}}} {\nu_{\mathrm{relax}}^{\mathrm{parabolic}}} = \frac{12}{\pi^2} = 1.216.
\end{equation}

The earlier sample--reservoir shear treatment assumed the sinusoidal profile for both the work and relaxation estimates; the present Eulerian profile analysis shows that the profile-dependent geometric factor for both work and relaxation is correctly the parabolic value.

\section{Cubic-spline Lennard-Jones potential}
\label{app:cubic-spline-lj}

The Lennard-Jones pair interaction was not suddenly truncated in our molecular-dynamics simulations, as is standard, but was instead matched to a cubic spline so that the potential and force go smoothly to zero at the cutoff, producing an intermediate-range interatomic interaction. This LJspline nevertheless reproduces the most important repulsive collisions and attractions for dense fluids and solids. Therefore, the cubic matching point was chosen as the inflection point of the LJ potential, where $\phi''=d^2\phi/dr^2=0$, i.e., $r_c=(13/7)^{1/6}=1.109$, where $\phi_c=\phi(r_c)=-0.787$ and $\phi_c'=\phi'(r_c)=2.689$. For $r\leq r_c$, the ordinary LJ potential Eq.~\ref{eq:lj} is used. For $r_c\leq r\leq r_m$, the LJ potential is replaced by a cubic spline:
\begin{equation}
\hat\phi(r) = \phi_c+\phi_c'(r-r_c)+a(r-r_c)^3,
\end{equation}
where the maximum cutoff, $r_m$, is chosen such that $\hat\phi(r_m)=0=\hat\phi'(r_m)$:
\begin{equation}
r_m-r_c=-\frac{3\phi_c}{2\phi_c'}, \qquad a=-\frac{\phi_c'}{3(r_m-r_c)^2}.
\end{equation}
For $r\geq r_m=1.548$, the interaction potential and force are identically zero.

In our simulations, we used a Verlet neighbor list with skin distance $r_{\mathrm{skin}}=0.3$ and list radius $r_{\mathrm{list}}=r_m+r_{\mathrm{skin}}$.

\printbibliography

\end{document}